\documentclass[11pt,final,onecolumn,letterpaper]{IEEEtran}

\usepackage[dvips]{graphicx}
\usepackage{cite}
\usepackage{graphicx}
\usepackage{amsmath}
\usepackage{amssymb}
\usepackage{multirow}
\usepackage{subfigure}
\usepackage{lscape}
\usepackage{etoolbox}
\makeatletter
\patchcmd{\@makecaption}
  {\scshape}{}{}{}
\makeatother

\graphicspath{{../Figs/}}
  
\newcommand{\tr}{\rule[-1ex]{0pt}{3.5ex} }
\newcommand{\muck}{$\mu_{C,K}$}
\newcommand{\mk}{$m^*_K$}

\newcommand{\app}{$\approx$}
\newcommand{\cmVs}{$\rm cm^2 V^{-1} s^{-1}$}

\begin{document}
\title{Experimentally Measured Field Effect Mobilities for Few Layer van der Waals Materials}

\author{Somaia~Sarwat~Sylvia and Roger~K.~Lake \\ 
University of California, Riverside, CA 92521 \\
January 12, 2016}
\maketitle
\begin{abstract}
A literature survey of experimentally measured mobility values for the two-dimensional
materials MoS$_2$, MoSe$_2$, MoTe$_2$, WS$_2$, WSe$_2$, and black phosphorous carried out during 2015.
\end{abstract}
\vspace{0.5in}

Tables \ref{tab:elec} and \ref{tab:hole} give experimentally measured mobility values for both electrons
and holes. The values for monolayer samples are highlighted in bold.
At the top of each column are the bandgaps.
At the bottom of each column,
are values of the monolayer effective mass at the valley indicated by the subscript.
The bandgaps and effective masses are from density functional theory calculations using the 
HSE hybrid functional \cite{Falko_allTMDC_bandstructure}.
The highest reported n-type and p-type mobility are both for WSe$_2$. 
For n-type WSe$_2$, the value is 202 cm$^2$/Vs, and for p-type WSe$_2$, the value is 250 cm$^2$/Vs.
Except for MoTe$_2$, WSe$_2$ has the lowest bandgap of 1.3 eV. 
MoTe$_2$ is the newest of the materials to be studied, so there is little experimental mobility data available.
All values are for room temperature unless otherwise noted.
All measurements are 2-point measurements unless otherwise noted.
Notes and references
for each value are provided after Table \ref{tab:hole}.
\vspace{3.5in}

\noindent
This work is supported in part by FAME, one of six centers of STARnet, 
a Semiconductor Research Corporation program sponsored by MARCO and DARPA.
The SRC publication number is P089463. 
\\
Email: ssylvia@ece.ucr.edu
\\
Email: rlake@ece.ucr.edu

\begin{table*}
\caption{
Electron mobilities (bold face numbers represent monolayer values). Theoretically calculated monolayer mobilities are also included for comparison. 
The superscripts refer to the number in the `Notes' section, which provides details about the sample and measurement. 
Source citations are given in the `Notes.'
}
\label{tab:elec}
\vspace*{-18pt}
\begin{center}       
\vspace*{8pt}
\begin{tabular}{|c|c|c|c|c|c|c|} 
\hline
\tr Material                  & MoS$_2$              & MoSe$_2$  & MoTe$_2$ & WS$_2$  & WSe$_2$      & BP  \\
\hline
\tr Bandgap (eV)          & 1.67                     &   1.40            &    0.997     & 1.60   & 1.30   & 1.55 \\
\hline
\tr Mobility                  &\textbf{130, \muck = 320}$^{\ref{allMobTh}}$
                                                                  &\textbf{25, \muck = 180}$^{\ref{allMobTh}}$
                                                                                       &\textbf{228}$^{\ref{allMobTh}}$
                                                                                                            &\textbf{320, \muck = 690}$^{\ref{allMobTh}}$
                                                                                                                           &\textbf{30}$^{\ref{allMobTh}}$
                                                                                                                                                      &$\mu_{C,x}$ = 1100$^{\ref{allMobTh}}$ \\
\tr (theoretical)           &                             &                    &                  &             &\textbf{\muck = 250}$^{\ref{allMobTh}}$
                                                                                                                                                      &$\mu_{C,y}$ = 80$^{\ref{allMobTh}}$\\                                                                                                                                                      
\tr \cmVs                     &&&&&& \\
\hline
\tr Mobility,                 &16.8$^{\ref{mos2.1}}$, \textbf{14.7}, 16.7$^{\ref{mos2.2}}$, 65.8$^{\ref{mos2.3}}$
                                                                   &\textbf{23}$^{\ref{mose2.1}}$              
                                                                                             & 0.9$^{\ref{mote2.1}}$, 40$^{\ref{mote2.1}, \ref{mote2.2}}$
                                                                                                            & 13.2, 34.7$^{\ref{ws2.1}}$    
                                                                                                                                  &\textbf{0.37}$^{\ref{wse2.5}}$ 
                                                                                                                                                   &275, 950, \textbf{94}$^{\ref{bp.7}}$\\
\tr $\rm{(cm^2/V-s)}$ &1.2$^{\ref{mos2.4}}$, \textbf{53}$^{\ref{mos2.5}}$, \textbf{44}$^{\ref{mos2.6}}$, \textbf{55}$^{\ref{mos2.7}}$
                                                                   &\textbf{50, 0.02}$^{\ref{mose2.2}}$ 
                                                                                             & 7$^{\ref{mote2.3}}$ 
                                                                                                             & 31$^{\ref{ws2.2}}$ 
                                                                                                                                  &\textbf{8.15, 3.67}$^{\ref{wse2.11}}$&78$^{\ref{bp.8}}$, 0.18$^{\ref{bp.10}}$\\
\tr                                &3.4$^{\ref{mos2.8}}$, 0.01-10, 46$^{\ref{mos2.9}}$
                                                                   & 2.14$^{\ref{mose2.3}}$ 
																					    & 30$^{\ref{mote2.4}}$ 
																					                     &\textbf{25, 49, 83}$^{\ref{ws2.3}}$ 
																					                                         &70$^{\ref{wse2.13}}$&40$^{\ref{bp.11}}$, 89$^{\ref{bp.13}}$, 27$^{\ref{bp.14}}$\\
\tr                                &69, 7, 18$^{\ref{mos2.10}}$
                                                                    & 35$^{\ref{mose2.4}}$ 
																						& 0.03$^{\ref{mote2.6}}$ 
																						                    &\textbf{4.1}$^{\ref{ws2.4}}$
																						                                    &30$^{\ref{wse2.14}}$&\app 85$^{\ref{bp.15}}$, 2 $^{\ref{bp.16}}$, 1, 0.5$^{\ref{bp.19}}$ \\
\tr                                &9.3$^{\ref{mos2.11}}$, 28$^{\ref{mos2.12}}$, 98.2$^{\ref{mos2.13}}$
                                                                    & 0.1 - 10$^{\ref{mose2.5}}$ 
                                                                                             &                  &\textbf{80, 163, 185}$^{\ref{ws2.5}}$
                                                                                                                                 &\textbf{7}$^{\ref{wse2.15}}$&12, 21.9, 38, 3$^{\ref{bp.19}}$ \\
\tr                                &22.4, 84$^{\ref{mos2.14}}$, \textbf{1-10}$^{\ref{mos2.15}}$
                                                                    & 43$^{\ref{mose2.6}}$ 
																						&                  &\textbf{4}, 30, 90, \textbf{15}$^{\ref{ws2.6}}$ 
																						                                    &\textbf{142, 202}$^{\ref{wse2.16}}$&0.12, 83$^{\ref{bp.19}}$ \\ 
\tr                                &7.4, 20.5, 25.2$^{\ref{mos2.17}}$
                                                                    & 19.7$^{\ref{mose2.7}}$ 
                                                                                             &                  &47, 125$^{\ref{ws2.6}}$, \textbf{0.91}$^{\ref{ws2.7}}$ && \\
\tr                                &\textbf{25.3, 10}$^{\ref{mos2.18}}$, \textbf{0.029}$^{\ref{mos2.19}}$, \textbf{7.24}$^{\ref{mos2.20}}$
                                                                    & \app 100 - 150$^{\ref{mose2.8}}$ 
                                                                                             &                  &60$^{\ref{ws2.8}}$ && \\
\tr                                &24$^{\ref{mos2.22}}$, 3.6, 8.2, \textbf{15.6}$^{\ref{mos2.23}}$                       
                                                                     &\app 50$^{\ref{mose2.8}, \ref{mose2.9}}$
                                                                                             &                  &\textbf{50}$^{\ref{ws2.9}}$ && \\
\tr                                &178, \textbf{1.7}, 8.1, 88.8$^{\ref{mos2.25}}$                     
                                                                     &                  &                  &12, 0.47$^{\ref{ws2.10}}$ && \\
\tr                                &\textbf{60}, 42, 36$^{\ref{mos2.26}}$
                                                                     &                  &                  &\textbf{0.23}, 17, 80$^{\ref{ws2.11}}$ && \\
\tr                                &3.1, 1.4, 2.1$^{\ref{mos2.27}}$
                                                                     &                  &                  &44, 19$^{\ref{ws2.12}}$ && \\
\tr                                &3.5, 12.5, 19, 46$^{\ref{mos2.28}}$
                                                                     &                  &                  &\textbf{10}$^{\ref{ws2.13}}$ && \\
\tr                                &0.9 - 7.6$^{\ref{mos2.29}}$  &&&&&\\   
\tr                                &30 - 60, 470$^{\ref{mos2.30}}$ &&&&&\\
\tr                                &91, 306.5$^{\ref{mos2.31}}$, \textbf{6, 16.1}$^{\ref{mos2.32}}$ &&&&&\\ 
\tr                                &184, 700$^{\ref{mos2.33}}$, \textbf{81}$^{\ref{mos2.34}}$ &&&&&\\
\hline
\tr  $m_e^*(\times m_o)$               & \mk =0.43     & \mk =0.49    & \mk =0.53    & \mk =0.35 & \mk = 0.39  & $m^*_{\Gamma,x}=0.17$ \\
\tr   &      &  &  &  & & $m^*_{\Gamma,y}=1.12$ \\
\hline
\end{tabular}

\end{center}
\end{table*}
%
%
\begin{table*}
\caption{
Hole mobilities (bold face numbers represent monolayer values). 
Theoretically calculated monolayer mobilities are also included for comparison. 
The superscripts refer to the number in the `Notes' section, which provides details about the sample and measurement. 
Source citations are given in the `Notes.'
}
\label{tab:hole}
\vspace*{-18pt}
\begin{center}       
\vspace*{8pt}
\begin{tabular}{|c|c|c|c|c|c|c|} 
\hline
\tr Material                             & MoS$_2$        & MoSe$_2$  & MoTe$_2$ & WS$_2$  & WSe$_2$     & BP  \\
\hline
\tr Bandgap (eV)                      & 1.67               &   1.40            &    0.997     & 1.60   & 1.30   & 1.55 \\
\hline
\tr Mobility                             &\textbf{270}$^{\ref{allMobTh}}$
                                                                        &\textbf{90}$^{\ref{allMobTh}}$
                                                                                             &$\times$
                                                                                                                &\textbf{540}$^{\ref{allMobTh}}$
                                                                                                                                  &\textbf{270}$^{\ref{allMobTh}}$
                                                                                                                                                      &$\mu_{V,x}$ = 640$^{\ref{allMobTh}}$ \\
\tr (theoretical)           &                             &                    &                  &             &                        &$\mu_{V,y}$ = 10,000$^{\ref{allMobTh}}$\\                                                                                                                                                      
\tr \cmVs                     &&&&&& \\
\hline
\tr Mobility,                             &2.4, 7.1$^{\ref{mos2.16}}$          
                                                                        &\textbf{15}$^{\ref{mose2.1}}$                      
                                                                                                & 26$^{\ref{mote2.3}}$          
                                                                                                                 &0.79$^{\ref{ws2.10}}$   
                                                                                                                                   &\textbf{1-100}$^{\ref{wse2.1}}$ 
                                                                                                                                                                           &413$^{\ref{bp.1}}$\\
\tr  $\rm{(cm^2/V-s)}$            &\textbf{2 - 12}$^{\ref{mos2.21}}$                       
                                                                        &\textbf{0.01}$^{\ref{mose2.2}}$ 
																					   & 10$^{\ref{mote2.4}}$ 
																					                        &43, 12$^{\ref{ws2.12}}$  
																					                                          &140$^{\ref{wse2.2}}$     &247, 392$^{\ref{bp.2}}$\\
\tr                                         &12.24$^{\ref{mos2.24}}$     
                                                                       & 10$^{\ref{mose2.3}}$ 
                                                                                            & 20 - 27$^{\ref{mote2.5}}$ 
                                                                                                                 &               &100$^{\ref{wse2.3},\ref{wse2.8}}$ 
                                                                                                                                                                         &184$^{\ref{bp.3}}$, 11.2$^{\ref{bp.4}}$\\
\tr                                         &480$^{\ref{mos2.30}}$
                                                                        &                & 0.3$^{\ref{mote2.6}}$ 
                                                                                                                 &               &\textbf{2, 66}$^{\ref{wse2.4}}$ 
                                                                                                                                                                         &17$^{\ref{bp.5}}$, 110, 78$^{\ref{bp.6}}$\\
\tr                                         &                          &                &                   &               &\textbf{5}$^{\ref{wse2.5}}$ 
                                                                                                                                                                        &400$^{\ref{bp.8}}$, 1350$^{\ref{bp.9}}$, 155$^{\ref{bp.10}}$\\                                                                                                                 
\tr                                         &                          &                &                   &               &318$^{\ref{wse2.6}}$   &63, 120$^{\ref{bp.11}}$, 44$^{\ref{bp.12}}$, 310$^{\ref{bp.13}}$\\                                                                                                                 
\tr                                         &                          &                &                   &               &350$^{\ref{wse2.7}, \ref{wse2.8}}$ 
                                                                                                                                                                        &200$^{\ref{bp.14}}$, 170$^{\ref{bp.15}}$, 240, 180$^{\ref{bp.16}}$\\      
\tr                                         &                          &                &                   &               &80$^{\ref{wse2.9}}$     &413$^{\ref{bp.17}}$, \textbf{116}$^{\ref{bp.18}}$, 984, 286$^{\ref{bp.19}}$\\
\tr                                         &                          &                &                   &               &\textbf{0.2}$^{\ref{wse2.10}}$ 
                                                                                                                                                                        &214, 205, 300, 100$^{\ref{bp.19}}$ \\                                                                                                                                                                                                                            
\tr                                         &                          &                &                   &               &\textbf{55, 96.6}$^{\ref{wse2.11}}$ 
                                                                                                                                                                        &170.5, 25.9, \app 200$^{\ref{bp.19}}$\\                                                                                                                                                                                                                            
\tr                                         &                          &                &                   &               &40$^{\ref{wse2.12}}$   &35, 400, 44, 172$^{\ref{bp.19}}$ \\
\tr                                         &                          &                &                   &               &180$^{\ref{wse2.14}}$ &49, 95.6, 310, 180$^{\ref{bp.19}}$ \\
\tr                                         &                          &                &                   &               &\textbf{90}$^{\ref{wse2.15}}$ 
                                                                                                                                                                        &600, 510, 25, 400$^{\ref{bp.19}}$ \\
\tr                                         &                          &                &                   &               &\textbf{250}$^{\ref{wse2.17}}$ & \\                                                                                                                                                                                                                            
\hline
\tr  $m_h^* (\times m_o)$                   & \mk = 0.54 & \mk = 0.64        & $\times$
                                                                                                                            & \mk = 0.35 & \mk = 0.36  & $m^*_{\Gamma,x}=0.15$\\
\tr   &  &  &                         & &  & $m^*_{\Gamma,y} = 6.35$ \\
\hline
\end{tabular}

\end{center}
\end{table*}

\newpage
Hole values next page.
\newpage
\begin{center}
Notes
\end{center}
\begin{enumerate}
\item \label{allMobTh} Theoretically predicted monolayer mobility. \muck ~provide the K-valley dominated electron mobility. No data was available for the valance band of MoTe$_2$ at the time of this work \cite{KWKim_PRB_TMDC_xport_prop, 14_TMDC_mobility, Nature_BP_mobility}.

\item \label{mos2.1} 13 nm MoS$_2$ FET with CYTOP passivation \cite{MoS2_13nm_nanotech}.
\item \label{mos2.2} 14.7 \cmVs~ is for monolayer and 16.7 \cmVs~ is for bilayer MoS$_2$ FETs \cite{MoS2_1L_apl2015}.
\item \label{mos2.3} 2.7 nm suspended (bridge channel) MoS$_2$ FET \cite{MoS2_3nm_susp_nanoscale}.
\item \label{mos2.4} 6-7 nm of MoS$_2$ mechanically exfoliated and then imprinted on substrate\cite{WSe2_6nm_appl_mat}.
\item \label{mos2.5} CVD grown monolayer back gated MoS$_2$ FET \cite{MoS2_1L_jap}.
\item \label{mos2.6} 4-point measurement on mechanically exfoliated monolayer back gated MoS$_2$ FET \cite{MoS2_1L_acsnano}.
\item \label{mos2.7} CVD grown monolayer top gated MoS$_2$ FET with high-k dielectric \cite{MoS2_1L_hfreq_nanolett}.
\item \label{mos2.8} Back-gate field-effect transistor based on 2-3 layers of MoS$_2$ thin film \cite{MoS2_2L_phot_res}.
\item \label{mos2.9} Mechanically exfoliated few layer (around 5 nm) MoS$_2$ back gated FET. For as-fabricated devices, mobilities range between 0.01 to 10 \cmVs~and is improved to 46 \cmVs~after deposition of ionic liquid droplet   \cite{MoS2_5nm_nanolett}.
\item \label{mos2.10} Trilayer h-BN (hexagonal Boron Nitride) encapsulated MoS$_2$ dual-gated FET with graphene contacts, two-terminal field-effect mobility is 69 \cmVs. By comparison, unencapsulated and HfO$_2$-encapsulated trilayer MoS$_2$ devices on SiO$_2$ with metal contacts showed lower mobilities of 7 and 18 \cmVs, respectively \cite{MoS2_3L_DG_acsnano}.
\item \label{mos2.11} Exfoliated bilayer MoS$_2$ transistor on BN substrate \cite{MoS2_2L_onBN_iop}.
\item \label{mos2.12} A uniform Al$_2$O$_3$ growth on MoS$_2$ basal plane by applying a remote O$_2$ plasma treatment prior to Al$_2$O$_3$ growth \cite{MoS2_12L_sci_rep}.
\item \label{mos2.13} 5 nm MoS$_2$ back-gated FETs. The average value of the extracted electron mobility for the transistors fabricated on pristine MoS$_2$ flakes is 37.6 $\pm$ 8.4 \cmVs and 27.5 $\pm$ 7.2 \cmVs~for transistors fabricated on hydrogen-treated MoS$_2$ flakes. On the other hand, the average mobility of the transistors fabricated on the sulfur-treated MoS$_2$ flakes is 85.9 $\pm$ 12.6 \cmVs. Transfer characteristics of a MoS$_2$ transistor that was first fabricated on a sulfur-treated MoS$_2$ flake and electrically tested once followed by a hydrogen post-treatment and electrical testing show  95.3 \cmVs~as-fabricated and 98.2 \cmVs~after the hydrogen post-treatment. \cite{MoS2_5nm_nanoscale}.
\item \label{mos2.14} 7 to 8 layers of mechanically exfoliated MoS$_2$. The mobility of pristine MoS$_2$ was measured as 22.4 \cmVs. After p-toluene sulfonic acid (PTSA) molecular doping exposure for 30 min,  mobility was improved  to 84 \cmVs \cite{MoS2_7L_iop}.
\item \label{mos2.15} Monolayer MoS$_2$ transistor with ultra high-k gate dielectric Pb(Zr$_0.52$Ti$_0.48$)O$_3$\cite{MoS2_1L_nanoscale}.
\item \label{mos2.16} CVD process using Mo(CO)$_6$ and H$_2$S as precursors, 3 layer back-gated MoS$_2$ FET show 2.4 \cmVs~and top-gated FETs show 7.1 \cmVs~\cite{MoS2_3L_nanoscale}.
\item \label{mos2.17} Mechanical exfoliation of 9.8 nm MoS$_2$. Mobility for MoS$_2$/SiO$_2$/Si and MoS$_2$/Al$_2$O$_3$/Si were 7.4 \cmVs~and 20.5 \cmVs, respectively and is further improved to 25.2 \cmVs~by HfO$_2$ passivation layer \cite{MoS2_10nm_aip_adv}.
\item \label{mos2.18} Monolayer MoS$_2$ top-gated FET with HfO$_2$ as gate oxide shows mobility 25.3 \cmVs~while back-gate devices (Si/SiO$_2$/MoS$_2$) show mobility 10 \cmVs~\cite{MoS2_1L_jjap}.
\item \label{mos2.19} 4 probe measurement on CVD grown monolayer MoS$_2$ \cite{MoS2_1L_jvstb}.
\item \label{mos2.20} CVD grown monolayer MoS$_2$ \cite{MoS2_1L_ieee_nanotech}.
\item \label{mos2.21} Monolayer MoS$_2$ via magnetron sputtering \cite{MoS2_1L_pfet_nanoscale}.
\item \label{mos2.22} Top-gated CVD MoS$_2$ FETs on Si$_3$N$_4$ with Al$_2$O$_3$ gate dielectric \cite{MoS2_1L_si3n4_apl}.
\item \label{mos2.23} 3.6 \cmVs~for monolayer, 8.2 \cmVs~for bilayer and 15.6 \cmVs~for trilayer MoS$_2$, CVD synthesized \cite{MoS2_1L2L3L_nanoscale}.
\item \label{mos2.24} The average value of field effect mobility  for 3.01 nm MoS$_2$ is 12.24 $\pm$ 0.741 \cmVs~ \cite{MoS2_3nm_pfet_apl}.
\item \label{mos2.25} 5.5 nm (8 layer) MoS$_2$ FETs are exposed to N$_2$ gas in the presence of deep ultraviolet light for different periods. Mobility varies between 178 \cmVs~to 220 \cmVs~depending on the exposure time \cite{MoS2_8L_rsc_adv}. For pristine mono-, bi- and few layer FETs, mobilities are 1.7, 8.1 and 88.8 \cmVs, respectively. These values are improved to \app 3.25, \app 13.5 and \app 136 \cmVs, respectively depending on exposure time \cite{MoS2_1L2LFL_duv_n2}.
\item \label{mos2.26} \emph{In situ} annealing on mono-, bi- and trilayer MoS$_2$. 4-probe mobility measurements give \app 60 \cmVs~for monolayer, 42 \cmVs~for bilayer and 36 \cmVs~for trilayer \cite{MoS2_1L2L3L_akis_nanoscale}.
\item \label{mos2.27} Top-gate MoS$_2$ FETs for bi- , tri-, and four- layer thin channels show mobilities of 3.1, 1.4, and 2.1 \cmVs, respectively \cite{MoS2_2L3L4L_nanoscale}.
\item \label{mos2.28} Bottom gated devices with 1-3 layers MoS$_2$, mobility is \app 50 \cmVs~for 1T phase electrode devices and \app 15 - 20 \cmVs~for the 2H contacts. For top gated devices, mobility is \app 12 - 15 \cmVs~for 1T metallic electrodes compared to 3–5 \cmVs~for Au on 2H electrodes. \cite{MoS2_nmat_phase_engg_contact}.
\item \label{mos2.29} Back gated mono and bilayer MoS$_2$ FETs \cite{MoS2_1L2L_nanolett}.
\item \label{mos2.30} MoS$_2$ field effect transistors on both SiO$_2$ and polymethyl methacrylate (PMMA) dielectrics,  mobilities are measured in a four-probe configuration. For multilayer MoS$_2$ on SiO$_2$, the mobility is 30 - 60 \cmVs, relatively independent of thickness (15 - 90 nm), and most devices exhibit unipolar n-type behavior. In contrast, multilayer MoS$_2$ on PMMA shows mobility increasing with thickness, up to 470 \cmVs~(electrons) and 480 \cmVs (holes) at thickness \app 50 nm \cite{MoS2_ambipolar_pmma_apl}.
\item \label{mos2.31} 12 nm MoS$_2$ based back-gated FETs. For a two-terminal configuration, mobility is 91 \cmVs~ which is considerably smaller than 306.5 \cmVs~ as extracted from the same device when using a four-terminal configuration \cite{MoS2_12nm_2pt_vs_4pt}.
\item \label{mos2.32} Monolayer MoS$_2$ FET, without a high-$\kappa$ top dielectric, mobility is 6.0 \cmVs~ which increased \app 3 times after Al$_2$O$_3$ overcoat \cite{MoS2_1L_electPerf}.
\item \label{mos2.33} 10 nm MoS$_2$, 184 \cmVs~ is without the high-$\kappa$ dielectric environment, 700 \cmVs~ is after the deposition of Al$_2$O$_3$ \cite{MoS2_appenzeller_nanolett}.
\item \label{mos2.34} Monolayer MoS$_2$, defects and interfaces are treated using thiol chemistry \cite{MoS2_1L_natcomm}.

\item \label{mose2.1} CVD grown, monolayer MoSe$_2$, measurements done for electric double layer transistors \cite{MoSe2_monolayer_acsnano2}.
\item \label{mose2.2} CVD grown, monolayer MoSe$_2$ \cite{MoSe2_monolayer_acsnano, MoSe2_monolayer_nanolett}.
\item \label{mose2.3} CVD grown, 20 nm thick channel of MoSe$_2$ \cite{MoSe2_TFT_scirep}.
\item \label{mose2.4} Mechanically exfoliated, 10 layers of MoSe$_2$\cite{MoSe2_10L_apl}.
\item \label{mose2.5} CVD grown, bilayer MoSe$_2$ \cite{MoSe2_5L_apl}.
\item \label{mose2.6} CVD grown, 5 layer MoSe$_2$ \cite{MoSe2_5L_apl}.
\item \label{mose2.7} 20 layers of MoSe$_2$, phototransistor structure \cite{MoSe2_25L_nanotech}.
\item \label{mose2.8} 5-15 nm thick MoSe$_2$, mechanicallly cleaved. Mobility is \app100 - \app150 \cmVs~ for parylene-C supported devices and \app50 \cmVs~ for SiO$_2$ supported devices \cite{MoSe2_5-15nm_acsnano}.
\item \label{mose2.9} 3-80 nm thick MoSe$_2$, micromechanical exfoliation, 4 point back gated device measurement \cite{MoSe2_3_80nm_apl}.

\item \label{mote2.1} Multi layered (\app 30 layers) phase patterned 2H-MoTe$_2$, 1T-MoTe$_2$. Mobility values are \app 0.9 \cmVs~ for 2H contacts, \app 40 \cmVs~ for 1T contacts \cite{MoTe2_science_phase_pattern}.
\item \label{mote2.2} 10 nm MoTe$_2$, synthesized using flux method \cite{MoTe2_10nm_nature}.
\item \label{mote2.3} \app 9 layer MoTe2, electrostatic gating using a solid polymer electrolyte, PEO:CsClO$_4$ \cite{MoTe2_9L_ion_gating_acsnano}.
\item \label{mote2.4} Surface transport using ionic liquid gated FETs. Carrier density was extracted from Hall effect measurements. The electron/hole mobility was calculated from the longitudinal conductivity \cite{MoTe2_surf_xport}.
\item \label{mote2.5} Mechanically exfoliated from CVD grown MoTe$_2$ single crystals. Mobilities range from \app20 \cmVs~ in bilayers to values approaching 30 \cmVs~ in seven layers \cite{MoTe2_2to7L_acsnano}.
\item \label{mote2.6} Mechanically exfoliated trilayer MoTe$_2$ FETs \cite{MoTe2_3L_adv_mat}.

\item \label{ws2.1} Mechanically exfoliated 5.2 nm WS$_2$ FET; 13.2 \cmVs~ is for pristine sample, 34.7 \cmVs~ is after LiF doping \cite{WS2_5nm_appl_mat}.
\item \label{ws2.2} Multilayerd (60 nm) WS$_2$ using pulse laser depositio (PLD) technique, mobility is measured by the Hall system \cite{WS2_60nm_nanoscale}.
\item \label{ws2.3} 4 probe measurements done on exfoliated monolayer WS$_2$ FETs. Mobility is 25 \cmVs~ with WS$_2$ on SiO$_2$, 49 \cmVs~ with Al$_2$O$_3$ in between WS$_2$ and SiO$_2$ and is improved to 83 \cmVs~ after MPS (3-mercaptopropyltrimethoxysilane) treatment \cite{WS2_1L_advmat}.
\item \label{ws2.4} Monolayer WS$_2$ was first synthesized on sapphire substrate using atmospheric pressure chemical vapor deposition method. To make the transistor device, WS$_2$ monolayers were delaminated and transferred onto SiO$_2$/Si substrates \cite{WS2_1L_acsnano}.
\item \label{ws2.5} The electron field-effect motilities of single layer (SL) WS$_2$ FETs on SiO$_2$, h-BN (hexagonal Boron Nitride) and h-BN/SL-WS$_2$/h-BN were 80, 163, and 185 \cmVs~ respectively. \cite{WS2_1L_BNsandwich}.
\item \label{ws2.6} Exfoliated single-, bi-, and five-layered WS$_2$ FETs were measured using deep-ultraviolet (DUV) light in nitrogen (N$_2$) and oxygen (O$_2$) gas environments. The field-effect mobilities of pristine monolayer, bilayer and five layer WS$_2$ FETs were 4, 30, and 90 \cmVs, respectively. After 30 min DUV + N$_2$ treatment, the field-effect mobilities improved to 15, 47, and 125 \cmVs~ \cite{WS2_1L2LML_UVexp}.
\item \label{ws2.7} CVD grown, monolayer WS$_2$ FET \cite{WS2_1L_nanoscale}.
\item \label{ws2.8} Chloride molecular doping on 5 - 7 monolayer WS$_2$ \cite{WS2_FL_Cldop_NL}.
\item \label{ws2.9} 4 terminal field effect mobility for monolayer WS$_2$ FET after 145 h of \emph{in situ} annealing\cite{WS2_1L_mobility_acsnano}.
\item \label{ws2.10} Multilayer WS$_2$ nanoflakes were exfoliated (42 nm thickness) from the commercially available WS$_2$ crystals  onto  SiO$_2$/Si substrates using conventional mechanical exfoliation technique. Mobility is 12 \cmVs. Few-layer WS$_2$ FETs exhibit ambipolar properties, the electron and hole mobility are calculated to be 0.47 and 0.79 \cmVs, respectively, \cite{WS2_FL_sci_rep}.
\item \label{ws2.11} Mechanically exfoliated flakes either on SiO$_2$ or h-BN/SiO$_2$ substrate. Mobility is 0.23 \cmVs~ for 1L (one layer)-WS$_2$/SiO$_2$, 17 \cmVs~ for 4L-WS$_2$/SiO$_2$ and \app 80 \cmVs~ for 4L-WS$_2$/h-BN/SiO$_2$ \cite{WS2_BN_sci_rep}.
\item \label{ws2.12} Mono- and bilayer WS$_2$ ionic liquid gated FETs, measurement temperature is 260 K. Mobility values are obtained as 44 (19) and 43 (12) \cmVs~ for electron and hole carriers in monolayer
(bilayer). \cite{WS2_1L2L_NL}.
\item \label{ws2.13} Micromechanical exfoliation of monolayer WS$_2$ \cite{WS2_1L_dalton_comm}.

\item \label{wse2.1} A series of back-gated FETs were fabricated on CVD monolayer WSe$_2$ crystals grown on Si$_3$N$_4$/SiO$_2$/Si substrates using graphene as source/drain electrodes, with Pd contact. FET mobilities  are in the range of 1 - 100 \cmVs~ \cite{WSe2_1L_adv_mat}.
\item \label{wse2.2} Dual-gated WSe$_2$ FET with Pt contacts underneath 3-4 layers of WSe$_2$, an h-BN top-gate dielectric, and a Pd top-gate. Hole mobilities are extracted using four-point measurements  \cite{WSe2_DG_3L_acsnano}.
\item \label{wse2.3} 6-7 nm of WSe$_2$ mechanically exfoliated and then imprinted on substrate\cite{WSe2_6nm_appl_mat}.
\item \label{wse2.4} Chemical vapor deposition (CVD) synthesized 2H phase monolayer WSe$_2$, the 1T phase is utilized as source and drain electrodes and intact CVD-grown 2H-WSe$_2$ as the channel material in FETs. The effective mobility of the 1T phase contacted WSe$_2$ devices is found to be \app 66.68 \cmVs, while it is \app 2 \cmVs~ with metal contacts.  \cite{WSe2_1L_phase_patt_acsnano}.
\item \label{wse2.5} CVD-grown WSe$_2$ monolayers on plastic substrates [poly(ethylene naphthalate), PEN]. Electric double layer transistors (EDLTs) were fabricated to evaluate the transport properties of these films. \cite{WSe2_1L_jjap}.
\item \label{wse2.6} PMMA encapsulated 5 layer WSe$_2$, gate stack is h-BN/HfO$_2$/Au \cite{WSe2_5L_apl}.
\item \label{wse2.7} 9-15 layers of WSe$_2$ \cite{WSe2_12L_sci_rep}.
\item \label{wse2.8} P-type WSe$_2$ field-effect transistors exhibit  hole carrier mobility up to 100 \cmVs~ for monolayer and up to 350 \cmVs~ for few-layer ($\geqslant$ 3L) materials\cite{WSe2_1L_FL_nanolett}.
\item \label{wse2.9} The backside of a 27 nm (40 layers) WSe$_2$ flake was encapsulated by fluorinated polymer (CYTOP) which allows to operate stably in air \cite{WSe2_27nm_apl}.
\item \label{wse2.10} Monolayer WSe$_2$ \cite{WSe2_1L_nanoscale}.
\item \label{wse2.11} Monolayer WSe$_2$, Au deposited, measurements done on EDLT. The highest carrier mobility of the hole transport increases from 55 to 96.6 \cmVs. After Au-doping, the electron mobility decreases from 8.15 to 3.67 \cmVs. \cite{WSe2_1L_2Dmat}.
\item \label{wse2.12} Sulfur-assisted CVD method to grow few-layer (5 nm/20 nm ) WSe2 flakes following a screw-dislocation-driven (SDD) growth fashion.\cite{WSe2_FL_acsnano}.
\item \label{wse2.13} 5 - 10 nm WSe$_2$ with a new n-doping scheme for TMDCs using PECVD SiN$_x$ \cite{WSe2_5to10nm_apl_mat}.
\item \label{wse2.14} Single- layer WSe$_2$ transistors gated by the polymer electrolyte PEO:LiClO$_4$. 4 point conductivity measurement, T = 250 K. \cite{WSe2_1L_akis_acsnano}.
\item \label{wse2.15} Ion gel dielectric, electric double-layer transistors (EDLT) of monolayer WSe$_2$ \cite{WSe2_1L_acsnano}.
\item \label{wse2.16} In-WSe$_2$ FET shows a high electron mobility of 142 \cmVs. By Al$_2$O$_3$ film deposition on WSe$_2$, the mobility of monolayer WSe$_2$ FET with Ag contact can reach around 202 \cmVs\cite{WSe2_DJena_Nano_Letters}.
\item \label{wse2.17} Top-gated monolayer transistors \cite{WSe2_pfet_Nano_Letters}.

\item \label{bp.1} 15 nm BP (Black Phosphorus) on Si/HfO$_2$ substrate \cite{BP_15nm_scirep}.
\item \label{bp.2} 14 to 28 nm BP FETs. 247 \cmVs~ is mobility in the zigzag direction and 392 is the same in the armchair direction \cite{BP_14to28nm_acsnano}.
\item \label{bp.3} 10 nm BP devices showed mobility of 184 \cmVs, but was shown to degrade with time \cite{BP_10nm_aip_adv}.
\item \label{bp.4} 5 nm BP FETs with HfO$_2$ encapsulation \cite{BP_5nm_apl}.
\item \label{bp.5} 10 nm top-gated BP FETs \cite{BP_10nm_shortCh_acsnano}.
\item \label{bp.6} Few layer BP FET, 110 \cmVs~ is along the zigzag direction and 78 \cmVs~ is along the armchair direction \cite{BP_FL_apl}.
\item \label{bp.7} 4-terminal measurement on few layer BP. 275 \cmVs~ is for 3 nm, 950 \cmVs~ is for 13 nm BP. 94 \cmVs~ is for monolayer BP, extracted using Matthiessen's rule \cite{BP_1L_extracted_mu_nat_comm}.
\item \label{bp.8} 10 nm BP FET on Si/SiO$_2$/h-BN (hexagonal Boron Nitride) substrate \cite{BP_10nm_onBN_nat}.
\item \label{bp.9} BN-(few layer)BP-BN heterostructure \cite{BP_FL_BNsandwch_natcomm}.
\item \label{bp.10} 5 nm BP with ferromagnetic tunnel contacts consisting of ultrathin layer of insulating TiO$_2$ and cobalt (Co) contacts. Hole and electron mobilities with four probe measurements are 150 \cmVs~ and 0.23 \cmVs, respectively \cite{BP_5nm_ferromag_contact}.
\item \label{bp.11} BP field effect transistors fabricated utilizing graphene as source drain electrodes and BN as an encapsulation layer. 63 and 120 \cmVs~ are mobilities with BP thickness of 4.5 nm and bilayer, respectively. 40 \cmVs~ is electron mobility for bilayer BP \cite{BP_2L_e_h_mobility_acsnano}.
\item \label{bp.12} 10.7 nm BP FET with 7 nm HfO$_2$ gate dielectric \cite{BP_10nm_EDL}.
\item \label{bp.13} Encapsulated bottom-gated 15 nm BP ambipolar FETs on flexible polyimide
\cite{BP_15nm_nanolett}.
\item \label{bp.14} 4.8 nm BP FET, Cs$_2$CO$_3$ is found to strongly electron dope black phosphorus. The electron mobility is significantly enhanced to \app 27 \cmVs~ after 10 nm Cs$_2$CO$_3$ modification. In contrast, MoO$_3$ decoration demonstrates a giant hole-doping effect although hole mobility is nearly maintained \cite{BP_8L_nat_comm}.
\item \label{bp.15} 10 nm BP FET on Si/SiO$_2$/h-BN substrate. Extracted room temperature mobilities from temperature vs mobility plot \cite{BP_10nm_apl}.
\item \label{bp.16} 240 \cmVs~ and 2 \cmVs~ are hole and electron mobilities for 15 nm BP. 180 \cmVs~ is hole mobility for 9 nm BP \cite{BP_9nm_15nm}.
\item \label{bp.17} 10 nm BP on HfO$_2$ \cite{BP_10nm_ieeeConfDSP}.
\item \label{bp.18} Monolayer BP FET \cite{BP_1L_sdas_nanolett}.
\item \label{bp.19} Table 3 of ref. \cite{BP_review_MatChemC}.
\end{enumerate}

\bibliographystyle{IEEEtran}

\end{document}